# JOINT ESTIMATION OF ANGLE AND DELAY OF RADIO WAVE ARRIVAL UNDER MULTIPLICATIVE NOISE ENVIRONMENT

*Pradip Sircar*

Department of Electrical Engineering
Indian Institute of Technology Kanpur
Kanpur 208016, India
<sircar@iitk.ac.in>

**SUMMARY**

We address the problem of joint estimation of angle and delay of radio wave arrival in a multipath mobile communication channel using knowledge of the transmitted pulse shape function [1-3]. Employing an array of sensors to sample the radio received signal, and subsequent array signal processing can provide the characterization of a high-rank channel in terms of the multipath angles of arrival and time delays [4]. An improved characterization of the channel can lead to better equalization process and effective transmit diversity [5]. Although several works have been reported in the literature for estimation of the high-rank channel parameters, we are not aware of any work that deals with the problem of estimation in a fading channel, which essentially leads to a multiplicative noise environment [6]. Indeed, it has been pointed out while developing the cumulant-based techniques for multiplicative noise environment that the technique is useless for the Rayleigh faded radio channels because the fourth-order cumulant of the received signal is identically zero in this case [7]. In the present work, we develop a unique technique for parameter estimation of a Rayleigh faded high-rank channel under the assumption of independent multipath rays [8]. Note that in [8], the problem considered is related to the low-rank channel model arising from local scattering around the mobile unit, and there is no distant scatterer / reflector present [4].

Consider an array of $M$ sensors receiving $L$ reflections of a known pulse shape function $g(t)$, where each propagation path is parameterized by the angle of arrival $\theta_i$ and the time delay $\tau_i$; and there is a complex reflection coefficient $\beta_i(t)$ associated with each path at time $t$. Under the assumption that the reflecting points are in the far field of the array, the output vector measured at the array of sensors is given by

$$\mathbf{x}(t) = \sum_{i=1}^{L} \beta_i(t)\, \mathbf{a}(\theta_i)\, g(t-\tau_i) + \mathbf{n}(t) \qquad (1)$$

where $\mathbf{a}(\theta_i)$ is the response vector of the array in the direction $\theta_i$, and $\mathbf{n}(t)$ is the noise vector [5]. The complex coefficient $\beta_i(t) = \gamma_i(t) e^{j\alpha_i(t)}$ is a random sequence such that the attenuation $\gamma_i$ is Rayleigh distributed and the phase $\alpha_i$ is uniformly distributed. It is assumed that the coefficient $\beta_i(t)$ is independent from snapshot to snapshot as well as from ray to ray. In fact, the method developed in this work under the condition of independence of $\beta_i(t)$ is valid when the attenuation is Rayleigh, Rician or Suzuki distributed [9]. Therefore, the developed method is equally applicable in various urban and suburban mobile communication scenarios.

For a uniform linear array (ULA) of element spacing $\delta$ in wavelength, the output signal at the $k$ th sensor neglecting noise is given by [5]

$$x_k(t) = \sum_{i=1}^{L} \gamma_i(t) e^{j\alpha_i(t)} e^{j2\pi\delta(k-1)\sin\theta_i} g(t-\tau_i) \qquad (2)$$

Taking the Fourier transform (FT) of both the sides, we get

$$\tilde{x}_k(\omega) = \sum_{i=1}^{L} \tilde{\gamma}_i(\omega) e^{j\tilde{\alpha}_i(\omega)} e^{j2\pi\delta(k-1)\sin\theta_i} e^{-j\omega\tau_i} \tilde{g}(\omega) \qquad (3)$$

where $\tilde{x}_k(\omega)$, $\tilde{g}(\omega)$, $\tilde{\gamma}_i(\omega)$ and $\tilde{\alpha}_i(\omega)$ are the FT 's of $x_k(t)$, $g(t)$, $\gamma_i(t)$ and $\alpha_i(t)$ respectively [10]. We compute the correlation between the frequency samples at the $k$ th and the $m$ th sensors as

$$c_{k,m} = E\{\tilde{x}_k(\omega)\tilde{x}_m^*(\omega)\} \qquad (4)$$

where $E$ stands for the expectation operator. On evaluation (4) reduces to

$$c_{k,m} = \sum_{i=1}^{L} G_i \, e^{j2\pi\delta(k-m)\sin\theta_i} \quad (5)$$
$$= c_\ell, \quad \ell = k - m$$

under the condition of independence of $\beta_i(t)$, where $G_i$ are real valued constants. It is clear from (5) that since the correlation $c_\ell$ is in the form of sum of exponential functions of $\sin\theta_i$, the estimates of $\sin\theta_i$ can be obtained by employing the singular value decomposition (SVD) based Prony's method [11].

Once the estimates of $\sin\theta_i = s_i$ are found, we compute the frequency samples $\xi_i(\omega)$ as follows

$$\xi_i(\omega) = \frac{1}{M} \sum_{k=1}^{M} e^{-j2\pi\delta(k-1)s_i} \tilde{x}_k(\omega) \quad (6)$$

which reduces to
$$|\xi_i(\omega)| = \tilde{\gamma}_i(\omega) |g(\omega)| \quad (7)$$
and
$$\angle \xi_i(\omega) = \tilde{\alpha}_i(\omega) - \omega\tau_i + \angle \tilde{g}(\omega) \quad (8)$$
for large $M$.

It can be seen from (8) that if $\tilde{\alpha}_i(\omega)$ is assumed to be zero mean, then the best-fit straight line of the frequency samples $\angle \xi_i(\omega) - \angle \tilde{g}_i(\omega)$ has the slope of $-\tau_i$. Thus, the estimates of $\tau_i$ can be obtained using the known samples of $\angle \tilde{g}(\omega)$.

For simulation study we use the raised-cosine pulse shape function $g(t)$ given by

$$g(t) = \frac{\sin(\pi t)}{\pi t} \, \frac{\cos(\pi \rho t)}{1 - 4\rho^2 t^2} \cos[2\pi f_c t + \pi.p(t)] \quad (9)$$

where $\rho$ is the excess bandwidth, $f_c$ is the carrier frequency and $p(t)$ is a random sequence of bits [12]. We set $\rho = 0.35$, $f_c = 0.25$, and collect samples over 32 symbol periods at the over sampling rate of 4. The pulse shape function $g(t)$ is plotted in Fig. 1. The magnitude spectrum of the function $g(t)$ is shown in Fig. 2, and the phase spectrum is shown in Fig. 3. After unwrapping the phase of the spectrum, we get the plot as shown in Fig. 4.

We choose the angles of arrival to be $-10°$ and $20°$, and the time delays are set at $-3$ and $-7$ respectively. An array of 64 sensors with element spacing half of wavelength are used for processing. The complex reflection (fading) coefficients are chosen to be complex Gaussian random variables with real and imaginary parts both having zero mean and unity variance. The estimated angles are found to be $-10.0000$ and $20.0001$ degrees. The best-fit straight lines are drawn in Figs. 5 and 6 using the frequency samples of $\xi_i(\omega)$. The slopes of the straight lines are computed for the estimates of $-\tau_i$. Some estimated time delays are shown in Table 1.

Table 1

| $-\tau_1$ | $-\tau_2$ |
|---|---|
| −3.2404 | −7.0124 |
| −3.0070 | −6.9589 |
| −3.2362 | −7.0295 |
| −3.2964 | −7.2644 |

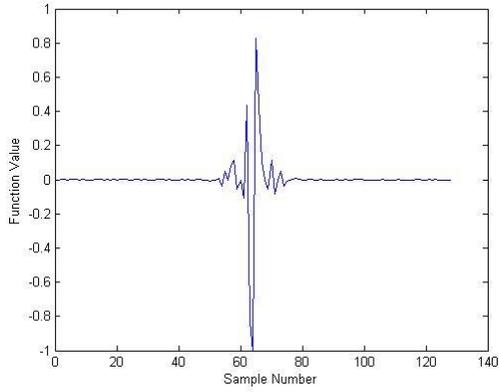

Fig. 1 The pulse shape function $g(t)$

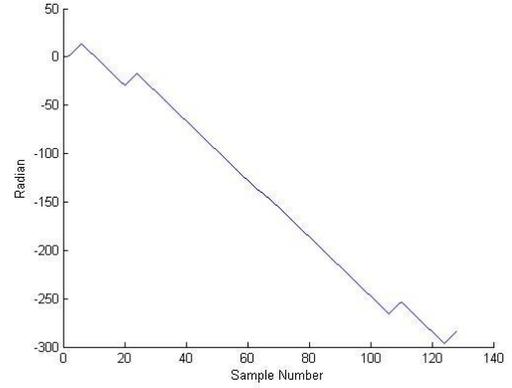

Fig. 4 The unwrapped phase spectrum of the function $g(t)$

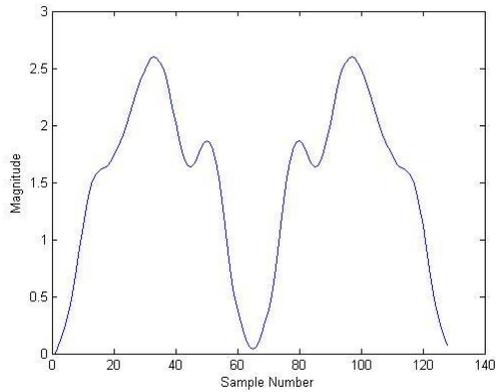

Fig. 2 The magnitude spectrum of the function $g(t)$

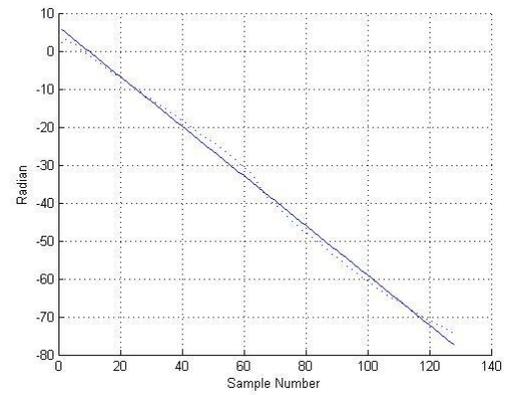

Fig. 5 The best-fit straight line of slope $-\tau_1$

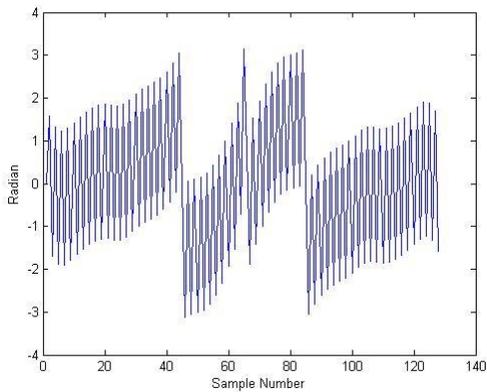

Fig. 3 The phase spectrum of the function $g(t)$

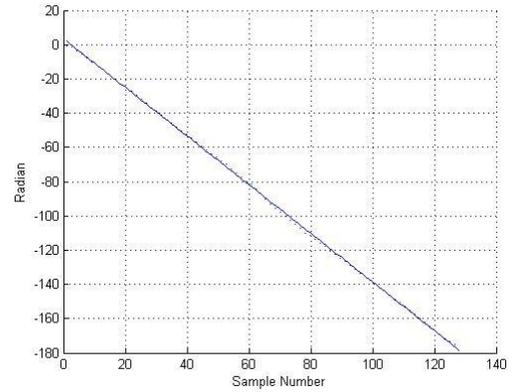

Fig. 6 The best-fit straight line of slope $-\tau_2$